\documentclass[%
 reprint,
showpacs,
 amsmath,amssymb,
 aps,
pra,
]{revtex4-1}

\usepackage{amsmath}
\usepackage{longtable}
\usepackage{graphicx}
\usepackage{dcolumn}
\usepackage{bm}
\usepackage{hyperref}
\usepackage[mathlines]{lineno}

\begin{document}

\preprint{APS/123-QED}

\title{New MC determination of the critical coupling in  $\phi^4_2$  theory.}

\author{Simone Bronzin$^{1}$}
\email{simone.bronzin@meta-liquid.com}

\author{Barbara De Palma$^{2}$}%
\email{barbara.depalma01@universitadipavia.it}

\author{Marco Guagnelli$^{1,2}$}
\email{marco.guagnelli@pv.infn.it}

\affiliation{
$^1$Via Antonini 20, Milano, Italy\\
$^2$ Universit\`a degli Studi di Pavia, Via A. Bassi 6, 27100, Pavia, Italy\\
$^3$ INFN, Sezione di Pavia, Via A. Bassi 6, 27100, Pavia, Italy\\
}%

\date{\today}

\begin{abstract}
	We investigate the non--perturbative features of $\phi^4$ theory in two dimensions, using Monte Carlo lattice methods.  In particular we determine the ratio $f_0\equiv g/\mu^2$, where $g$ is the unrenormalised coupling, in the infinite volume and continuum limit. Our final result is $f_0= 11.055(14)$.
\pacs{12.38.Gc, 11.15Ha}
\end{abstract}

\keywords{Suggested keywords}
\maketitle


\section*{Introduction}
$\phi^4$ theory plays an important role in Quantum Field Theory as it represents for example an extremely simplified model for the Higgs sector of the Standard Model. 

In $D=2$ dimensions the theory is super--renormalizable: the coupling constant $g$ has positive mass dimensions $[g]=[\mu_0^2]$, where $\mu_0$ is the (bare) mass parameter of the theory; this means that the ratio $f \equiv g/\mu^2$, where $\mu^2$ is a renormalised squared mass in some given renormalisation scheme, is the only physically relevant dimensionless parameter we have to consider. Thanks to the super--renormalisability of the theory we can use the unrenormalised coupling constant in the definition of $f$, since in any case the renormalisation of $g$ amounts to a finite constant.


In this paper we determine the value of $f \equiv g/\mu^2$ at the critical point, that is the value of $f$ computed in the limit in which both $g$ and $\mu^2$ go to zero. We follow the renormalisation scheme used in~\cite{ref7:LoinazWilley,ref6:SchaichLoinaz}, adopting the simulation technique introduced in~\cite{ref19:WolffConf}, namely the \emph{worm algorithm} and we compute the ratio $g/\mu^2$ using the same strategy implemented in~\cite{ref:paper8}; we present an improvement in the determination of the critical value $f_0$, obtained thanks to the \emph{gradient flow}~\cite{wf:monahan2015}, a technique that allows us to reach smaller values of the coupling $g$ with respect to our previous work~\cite{ref:paper8}.

In the following, after briefly describing the model and the renormalisation scheme chosen in order to extract $\mu^2$ at fixed $g$ in the infinite volume limit, we will recall the main steps of the simulations, focusing on the application of the gradient flow; we will then proceed to the continuum limit extrapolation. In the end we will compare our results with our previous determination of the same quantity and we will draw some conclusions. \\

\section{Lattice formulation}

Let's introduce the $\phi^4$ Lagrangian in the Euclidean space:
\begin{equation}
\mathcal{L}_E = \dfrac{1}{2}\left(\partial_\nu\phi \right)^2 + \dfrac{1}{2}\mu_0^2\phi^2 + \dfrac{g}{4}\phi^4.
\end{equation}
In $D=2$ the Euclidean action is
\begin{equation}
\mathcal{S}_E = \int d^2x\,\mathcal{L}_E.
\end{equation}
In order to obtain a dimensionless discretized action we put the system on a 2-dimensional lattice with spacing $a$ and linear size $L \equiv Na$. By introducing the following parametrization
\begin{equation}
\label{ga-mua}
\hat{\mu}^2_0=a^2\mu^2_0, \qquad \hat{g}=a^2g.
\end{equation}
we have
\begin{equation}
\mathcal{S}_E =  \sum_x \left\lbrace -\sum_{\nu}\phi_x\phi_{x+\hat{\nu}} + \dfrac{1}{2}\left( \hat{\mu}_0^2+4\right)\phi_x^2 + \dfrac{\hat{g}}{4}\phi_x^4 \right\rbrace, 
\end{equation}
where $\phi_{x\pm\hat{\nu}}$ are fields at neighbor sites in the $\pm\nu$ directions. 

In the following we will omit the ``hat'' on top of lattice parameters: all quantities will be expressed in lattice units, {\em i.e.} they become dimensionful when multiplied by appropriate powers of the lattice spacing $a$.

Numerical simulations are necessarily performed at finite values of $a$, $L$ and of the bare parameters of the theory: to obtain $f_0$ we have to extrapolate our results to $L \to \infty$ (infinite volume or thermodynamic limit) and to $a \to 0$ (continuum limit).


In order to approach the thermodynamic limit we fix $g$ to a given value in lattice units, and this amounts to keep fixed the lattice spacing $a$; we then simulate the system at several values of $N = L/a$. For each value of $N$ we perform several simulations searching for a value of $\mu^2_0$ such that a certain condition is satisfied. This condition, which we will describe in details later on, is conceived in such a way that by extrapolating $\mu^2_0(g,L/a)$ to the infinite volume limit we get a second order phase transition point in the plane $(g,\mu^2_0)$.

As we discuss in~\cite{ref:paper8}, in order to safely go to the continuum limit, we have to work out a renormalisation of the mass parameter, since $\mu_0^2$ in this limit diverges like $\log(a)$. Adhering to the same renormalization procedure adopted in~\cite{ref7:LoinazWilley,ref6:SchaichLoinaz},   
we determine the renormalised squared mass $\mu^2$ putting it equal to the solution, in the infinite volume limit, of the equation
 \begin{equation}
 \mu^2 = \mu^2_0 + 3gA(\mu^2).
 \label{eq:defmu2}
 \end{equation}
where $A(\mu^2)$ is the only 1--Particle--Irreducible divergent diagram in $D=2$ on a $N\times N$ lattice:
\begin{equation}
A(\mu^2_0) = \dfrac{1}{(2\pi)^2} \int \dfrac{\textrm{d}^2 p}{p^2 + \mu^2_0}
\end{equation}

	
The condition~\eqref{eq:defmu2} is equivalent to the introduction of a proper divergent mass--squared counterterm in the action. 
We may finally extrapolate the quantity $f\equiv g/\mu^2$ to $g\to 0$ in order to obtain $f_0$, the critical value in the continuum limit. 

Another useful parametrization of the action is the following:
\begin{equation}\label{eq:sbetalambda}
\begin{split}
\mathcal{S}_E &= -\beta\sum_x\sum_\nu\varphi_x\varphi_{x+\hat{\nu}} + \sum_x\left[\varphi^2_x + \lambda(\varphi^2_x-1)^2\right] \\
&= \mathcal{S}_{I} + \mathcal{S}_{Site},
\end{split}
\end{equation}
where the relations between $(\mu_0^2,\;g)$ and $(\beta,\;\lambda)$ are:
\begin{equation}\label{param}
\phi_x = \sqrt{\beta}\varphi, \qquad \mu_0^2 = 2\dfrac{1-2\lambda}{\beta}-4, \qquad g=\dfrac{4\lambda}{\beta^2}.
\end{equation} 

In eq.(\ref{eq:sbetalambda}) there is an interaction term between neighbor sites, $\mathcal{S}_{I}$, with a coupling constant of strength $\beta$ and a term related to a single site, $\mathcal{S}_{Site}$. 

\subsection{Simulations}
Now we outline the general computational strategy, focusing in particular on the improvements with respect our previous work ~\cite{ref:paper8}. 

In our simulations we used the \emph{worm algorithm}~\cite{ref19:WolffConf} and used the lattice action given by \eqref{eq:sbetalambda}.
Operatively we fix a value of $\lambda$ and $L/a$ and search for a value of $\beta$ such that the physical condition
\begin{equation}\label{mL}
mL = L/\xi = \textrm{const} = z.
\end{equation}
is matched for a given and fixed value of $z$. Condition~\eqref{mL}
implies that the correlation length $\xi$ of the system grows linearly with $L$: thus, when $a/L\to 0$, we arrive at the critical point, where the correlation length $\xi$ diverges if measured in units of the lattice spacing.
We then perform several simulations using different values of $N\equiv L/a$; for each couple $(\lambda,N)$ we obtain a particular value of $\beta_c(\lambda,N)$ such that the condition~\eqref{mL} is satisfied. $\beta_c(\lambda)$ is then obtained by extrapolating our results to $a/L\to 0$. As explained and numerically demonstrated in \cite{ref:paper8}, Renormalisation Group arguments ensure us that for small enough values of $a/L$ we can extrapolate $\beta_c(\lambda,a/L)$ linearly in $a/L$.

Using the relations in~\eqref{param} we compute $g(\lambda,\beta_c)$ and $\mu_0^2(\lambda,\beta_c)$. Then, using the renormalization condition~\eqref{eq:defmu2}, we get the value of $\mu^2(g)$  and hence the ratio $f\equiv g/\mu^2$. This procedure is repeated for several values of $\lambda$ (and hence of $g$) and finally, in order to obtain $f_0$, we extrapolate our results to $g\to 0$.

We now focus on the condition~\eqref{mL}. In this work we introduce a slight modification in the procedure for the computation of the mass parameter $m$: it is implicitly defined by the condition
\begin{equation}\label{eq:defm}
R_\rho \equiv \dfrac{G_\rho(\tau, p^*)}{G_\rho(\tau, 0)} = \dfrac{m^2}{p^{*2} + m^2}, 
\end{equation}
where $p^*$ is the smallest momentum on the lattice.
We decide to compute the propagator $G_\rho(\tau,p)$ using the gradient flow technique in the contest of scalar field theory (see for example~\cite{wf:monahan2015}). In particular, considering the action \eqref{eq:sbetalambda}, we introduce a new scalar field $\rho(x,\tau)$ depending on the space--time index $x$ and on the so--called flow--time $\tau$. The flow--time evolution equation of  $\rho(x,\tau)$ is 
\begin{equation}
\dfrac{\partial}{\partial \tau} \rho(x,\tau) = \partial^2\rho(x,\tau)
\end{equation}
where $\partial^2$ is the Laplace operator acting in the configurations space. If we now impose the Dirichlet boundary conditions, that is $\rho(x,0) = \varphi_x$, it is easy to write the exact solution, in the momentum space, for the propagator of the field $\rho$ at flow--time $\tau$:
\begin{equation}
G_\rho(\tau,\rho) = e^{-2\tau p^2}G(p).
\end{equation}
In this way we obtain a smearing effect of the original fields, since the flow--time exponentially suppresses the ultraviolet modes. To the total flow--time $\tau$ we can associate a smearing radius $r_{sm}=\sqrt{2d\tau}$, where $d$ is the dimensionality of the space--time. For a certain value of $\tau$, the ultraviolet suppression effect of the flow--time helps us to obtain values which are closer to the continuum limit. In this way, at fixed $\lambda$, we expect to safely extrapolate $\beta_c(\lambda,a/L) \to \beta_c(\lambda,0)$ using not too large values of $L/a$ with a consequently reduction of both the computational time and the statistical errors.\\

\begin{figure}[h]
	\centering
	\includegraphics[width=0.5\textwidth]{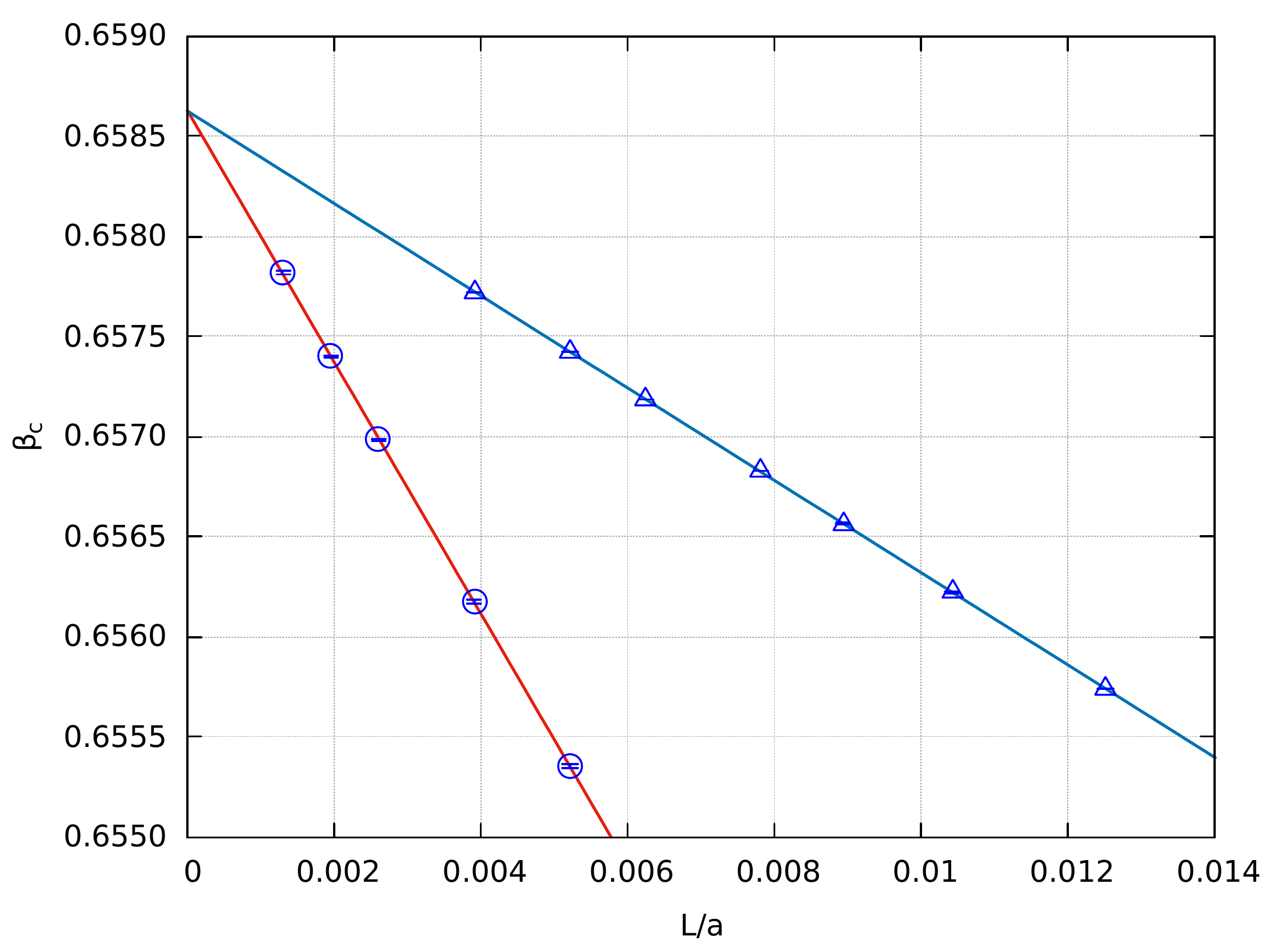} 
	\caption{Comparison between our previous results(lower set of points) and the new ones at $\lambda = 0.25$.}\label{fig:zold-zwf}
\end{figure}

Some preliminary simulations convinced us to assume the condition $z_\rho= mL = 1$ and to fix the value of $\tau$ such that at different $a/L$ values the smearing radius is equal to $L/4$. In this way we take advantage of the smearing effect of the gradient flow and take under control finite volume effects. In Fig.~\ref{fig:zold-zwf} we show the extrapolation of $\beta_c(L)$ at $\lambda=0.25$: the blue line is the extrapolation obtained by fixing  $z_\rho=1$, while the red line is the one obtained fixing $z=4$ and that we used in our previous work~\cite{ref:paper8}. As one can see, the new results (triangular points) show a linear behaviour even for small lattice sizes.

At fixed value of $\lambda$ we simulate the system for several values of $L/a$, namely: $L/a=$ 32, 40, 48, 56, 72, 80, 96, 112, 128, 144, 192, 256, 320, 384, 448, 512. We perform our simulations for the coupling values $\lambda =$ 0.005, 0.004, 0.003, 0.002, 0.001, 0.00075, 0.0005. The number of thermalisation  sweeps for all our simulations is several hundreds times the autocorrelation time of~\eqref{mL}. We keep under control the autocorrelation time of our observables using a Python program described in~\cite{Alg:unew}, based on~\cite{ref:LessErr}. We perform 1000 worm--sweeps between two consecutive measurements  and the numbers of measurements varies from $10^4\div 10^5$, according to $L$ and $\lambda$.

\section{Results}

\begin{table}[h!]
	\caption{\label{tab:wf-result} Infinite volume results of $\beta_c$ and $f(\lambda)$ with different linear lattice sizes.}
	\centering
	\begin{tabular}{cccccc}
		\hline
		\hline
		$\lambda$  &   $\beta_c$     & fit type   & $(L/a)_{\textrm{min}}$  & $(L/a)_{\textrm{max}}$  & $f(\lambda)$  \\
		\hline
		$0.0005$   & $0.5019535(5)$  & linear     & $192$                   & $512$                   & $10.9920(88)$  \\
		$0.00075$  & $0.5027800(10)$  & linear     & $144$                   & $384$                   & $10.9754(79)$  \\
		$0.001$    & $0.5035613(10)$ & linear     & $112$                   & $384$                   & $10.9258(79)$  \\
		$0.0015$    & $0.5050340(2)$ & linear     & $112$                    & $384$                   & $10.8801(20)$ \\
		$0.002$    & $0.5064156(9)$ & linear     & $112$                    & $384$                   & $10.8304(45)$ \\
		$0.003$    & $0.5089871(7)$ & linear     & $128$                   & $384$                   & $10.7504(15)$  \\
		$0.004$    & $0.5113712(4)$ & linear     & $128$                   & $384$                   & $10.6884(15)$ \\
		$0.005$    & $0.5136155(5)$  & linear     & $112$                   & $512$                   & $10.6435(10)$ \\
		\hline
		\hline
	\end{tabular} 
\end{table}
In Tab. \ref{tab:wf-result}
we report the infinite volume results at the several $\lambda$ we simulate.  These results are obtained performing a linear extrapolation in which the smallest values of $L$ are excluded. All the extrapolations give a final $\chi^2/d.o.f\sim1$. 
We finally extrapolate $f(\lambda)\to f_0$ as $\lambda\to 0$ by using both a linear fit functions (using only the last four points) and a quadratic fit function using all points.   
Our final results are 
\begin{align}
\label{eq:res2}
f_0^{quad}&= 11.058(4) &\chi^2=0.71 \qquad\text{with } 5 \text{ d.o.f}, \\
f_0^{lin\,}&=11.053(13) &\chi^2=1.79 \qquad\text{with } 2 \text{ d.o.f}.
\label{eq:res3}
\end{align}

We take the almost perfect agreement of these two results as a numerical evidence of the fact that a simple polynomial law is enough to describe the behaviour of $f(\lambda)$ near $\lambda\to 0$.

We decide to take the mean value of the two results as our final value and the difference between~\eqref{eq:res2} and ~\eqref{eq:res3} as an estimate of the systematic error involved in the extrapolation, adding it in quadrature to the statistical error. We finally quote
\begin{equation}
f_0 = 11.055(14)
\end{equation}
to be compared to our previous result:
\begin{equation}
\label{eq:res1}
f_0= 	11.15(6)(3)
\end{equation}
Considering both the statistical and the systematic error in~\eqref{eq:res1}, the two values are well compatible within 2--sigma level but the new result~\eqref{eq:res2} has a sensibly reduced error.
\begin{figure}[h]
	\centering
	\includegraphics[width=0.5\textwidth]{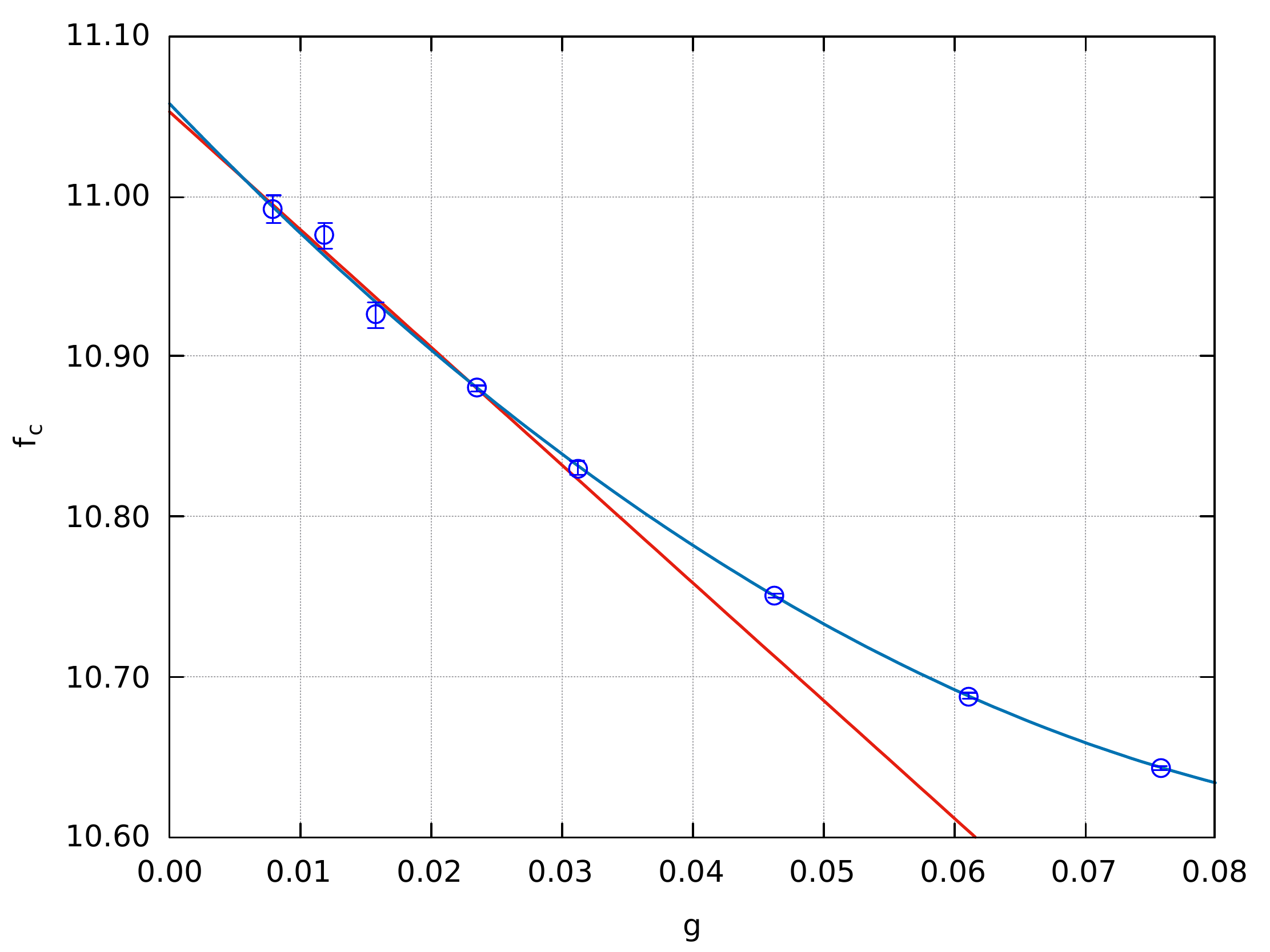} 
	\caption{Final extrapolation of $f$ versus $g$. }\label{fig:div-diagrams}
\end{figure} 

\section{Conclusions}
In table~\ref{tab:review-2} we summarize some of the latest results of $f_0$ derived with different approaches: the works \cite{ref14:DLCQ,ref15:QSE,ref16:DMRG,ref17:Umatrix,Elias-Miro:2017tup,Elias-Miro:2017xxf} are based on Hamiltonian truncation (variational) methods, Borel summability is applied  in~\cite{Serone:2018gjo,ref:vicari-peli}, where in~\cite{ref:vicari-peli} lattice results are used. Finally in~\cite{ref20:SLAC} lattice theory is simulated by using non--local SLAC derivative. Since we are in a good agreement with our previous result, the same considerations are still valid: our result is compatible with the last six determinations (excluding~\cite{ref:burkardt,ref:paper8}) at the 2$\sigma$--level, which come from different methods. 

The gradient flow technique allows us to reach lower values of $\lambda$ with respect to our previous work and to obtain a more precise estimation of $f_0$.

\begin{table}[h]
	\caption[Latest determination of the critical coupling in $\phi^4_2$ theory plus our final results]{\label{tab:review-2}Sample of the results for the continuum critical parameter $f_0$ from the literature. DLCQ stands for \emph{Discretized Light Cone Quantization}, QSE diagonalization for \emph{Quasi--Sparse Eigenvector} diagonalization, DMRG for \emph{Density Matrix Renormalization Group} and for  DLCH-FS \emph{Diagonalized light-front Hamiltonian in Fock-Space representation}.}
	\centering	
	\begin{ruledtabular}
		\begin{tabular}{lcc}
			Method						& $f_0$					&		year, Ref.	\\
			\hline
			DLCQ						& $5.52$              	&	1988, \cite{ref14:DLCQ}			\\
			QSE diagonalization			& $10$                	&	2000, \cite{ref15:QSE}			\\
			DMRG						& $9.9816(16)$     		&	2004, \cite{ref16:DMRG}			\\
			Monte Carlo cluster			& $10.8^{0.1}_{0.05}$ 	&	2009, \cite{ref6:SchaichLoinaz}	\\
			Monte Carlo SLAC derivative	& $10.92(13)$			&	2012, \cite{ref20:SLAC}			\\
			Uniform Matrix product states     & $11.064(20)$        	&	2013, \cite{ref17:Umatrix}		\\
			Monte Carlo worm			& $11.15(6)(3)$       	&	2015, \cite{ref:paper8}	\\
			Borel summability--Lattice Results 			& $11.00(4)$			&   2015, \cite{ref:vicari-peli}\\	
			DLCH-FS						& $4.40(12)$			&	 2016, \cite{ref:burkardt}\\
			Renormalised Hamiltonian	& $11.04(12)$	&	2017, \cite{Elias-Miro:2017tup,Elias-Miro:2017xxf}\\
			Borel summability			& $11.23(14)$	&	2018, \cite{Serone:2018gjo} \\
			This work & $11.055(14)$  & 2018
		\end{tabular} 
	\end{ruledtabular}
\end{table}


\bibliographystyle{apsrev4-1}
\bibliography{bib}

\end{document}